\documentclass[11pt,twoside]{article}

\title{\textbf{On the implicite interest rate in the Yunus equation}}
\author{Marc Diener, Pheakdei Mauk}
\date{}
\usepackage{latexsym}  
\usepackage[dvips]{graphicx}

\newtheorem{theoreme}{Th\'eor\`eme}

\newtheorem{proposition}[theoreme]{Proposition}

\newenvironment{exemple}{\pagebreak[1]\par\medskip {\noindent \bf Exemple~:\ }}{\medskip}

\newcommand{\ovl}{\overline}

\newcommand{\enreserve}[1]{}
\newcommand{\nop}{}

\newcommand{\zerobar}{\mbox{o\hspace{-0.5em}/}}

\input amssym.def
\input amssym
\let\BBB\Bbb
\def\Bbb#1{\ifmmode{\BBB #1}\else{${\BBB #1}$}\fi}

\begin{document}
\maketitle
\newcommand{\binom}[2]
             {\left(\begin{array}{c}
                           {#1}\\{#2} 
                    \end{array}
             \right)}     
\section{Introduction~: microcredit}
Microcredit is a set of contracts taylored to provide very small loans to very 
poor people to help develop small businesses or activities
generating income. The basic idea came from the finding that a large
part of humanity has no access to traditional credit because banks require 
their borrowers to meet a range of criteria, such as being able to read and
write, bears some identification documents, or to have already secured a 
minimum deposit.
The first experiments date back to the 70s in Bangladesh as an initiative of 
Muhammad Yunus, then a professor of economics at Chittagong University. In 
1974, he watched helplessly as a terrible famine in the little village 
 Joha near to his University. He then with 
his students asks the craftsmen and peasants of the 
village in order to 
try to understand their needs and lists a demand for 
small loans for 42 
women to whom he finally decides to pay himself a total of about 27 Euros. 
Then he spends nearly 10 years trying to persuade banks to take on these loans 
before
finally deciding to start his own bank, the Grameen Bank in 1983. This
bank and himself receive the Nobel Prize for Peace in 2006. 
Currently microcredit activity has spread to most countries in the world, it 
is ensured by close
10 000 Micro Finance Institutes (MFIs) who lend 50 billions euros to almost
500 millions beneficiaries.
The main characteristics of microcredit are
\begin{itemize}
\item Very small loans over short periods (10 Euros a year) with
 frequent (weekly) reimbursements.
\item Beneficiaries are mostly women.
\item Borrowers can't provide personal wealth to secure the loan.
\item Usually loans are with joint liability of a group of borrowers
(5 to 30) each borrower receiving her loan individually but all are 
interdependent in that they
must assume all or part of the failure (called the ``default") of any member 
of the group.
\item Interest rates are high, around 20\%, some of them up to 30\%.
\item Possibility of a new loan granted automatically in case of timely 
refunds (dynamic incentive mechanism).
\item repayment rate close to 100\%.
\end{itemize}

\section{The Yunus polynomial and equation}
\begin{exemple}
The following example has been given by Muhammad Yunus
\cite{Yunus}\cite{Yunus1}.



Grameen lends 1000 BDT (Bangladesh Taka) 
to borrowers that pay back 22 BDT%
\footnote{The value of $100$ Bangladesh Taka (BDT) is about  $1$ Euro.}
 each week during 50 weeks.
Let's denote by $r$ the annual continuously compound interest rate.
The present value of the $22$ BDT refunded after one week 
is $22e^{\frac{-r}{52}}$ 
this value of those of the next payment is 
$22e^{\frac{-2r}{52}}$ \ldots and so on. 
So, letting 
$q=e^{-\frac{r}{52}}$, as the 50 refundings balance the 1000 BDT received, we 
get following equation for $q$~:
\begin{equation}\label{Yunusdeterministe}
1000=22\sum_{k=1}^{50}q^k=22\frac{q-q^{51}}{1-q}
\end{equation}
that reduces to (the degree 51 polynomial equation) ${\cal{Y}}(q)=0$ where ${\cal{Y}}$ 
denotes what we shall call que {\em Yunus polynomial}\nop 
\begin{equation}
\label{Yunuspolynomial}
{\cal{Y}}(q):=22q^{51}-1022q+1000.
\end{equation}
We observe that ${\cal{Y}}$ has obviously $q=1$ as zero, has two other zeros 
$q_-<0<q_+<1$, and all other zeros are complex conjugate. An approximation of 
$q_+$ gives $q_+=0.9962107\ldots$ which leads to $r=19,74\ldots$, so nearly  $20\%$.
\end{exemple}

But some borrowers don't pay in time, so the $n$-th payment 
takes place
at some random time 
$T_n=T_{n-1}+\frac{1}{52}X_n=\frac{1}{52}(X_1+X_2+\ldots+X_n)$

Lets assume $(X_i)_{i=1..50}$ i.i.d, 
$X_i\leadsto {\cal{G}}(p)$, the geometric distribution, $p=\mbox{\Bbb{P}}(X_i=1)$, close to $1$~; in 
other words each week the borrower has probability $p$ to be able to pay the 
$22$ BDT she should pay, weekly refunding accidents being assumed to be 
independent.
So $r$ becomes a random variable, $R=r(X_1,\ldots,X_{50})$, satisfying the 
``Yunus equation"~:
\begin{equation}\label{Yunusequation}
1000=\sum_{n=1}^{50}22e^{-\frac{R}{52}(X_1+\ldots+X_n)}
\left(=22\sum_{n=1}^{50}v^{R(X_1+\ldots+X_n)}\;\;,\;\;\;\; v=e^{-\frac{1}{52}}\right).
\end{equation}
For the sake of getting a better understanding of the risks faced by the 
lender under these new asumptions we wish to have informations on the 
probability law of the random variable $R$. The sequel of this paper is 
devoted to the results we got so far.

\section{Actuarial expected rate}
Let us call {\em actuarial expected rate }the positive real number $\ovl r$ 
such that, replacing $R$ by 
$\ovl r$, it satisfies the expectation of the Yunus equation~:
\begin{eqnarray*}\label{EYunus}
1000&=&\mbox{\Bbb{E}}\left(\sum_{n=1}^{50}22v^{-\ovl{r}(X_1+\ldots+X_n)}\right)\;\;,\;\;\;\;v=e^{-\frac{1}{52}}\\
&=&22\sum_{n=1}^{50}\mbox{\Bbb{E}}\left(v^{-\ovl{r}X_1}\right)\ldots\mbox{\Bbb{E}}\left(v^{-\ovl{r}X_n}\right)
\mbox{ , as $X_1$\ldots$X_n$ are independent }\\
&=&22\sum_{n=1}^{50}\ovl{q}^n=22\;\;\frac{\ovl q-\ovl{q}^{51}}{1-\ovl q}
\mbox{ , with }\ovl q=\mbox{\Bbb{E}}(e^{-\frac{\ovl{r}}{52}X_1})
=M_{X_1}\left(-\frac{\ovl{r}}{52}\right),
\end{eqnarray*}
where $M_{X_1}(t)=\frac{pe^t}{1-(1-p)e^t}$ is the moment generating function of 
${\cal{G}}(p)$. So $M_{X_1}\left(-\frac{\ovl{r}}{52}\right)=\ovl q=q_+$, 
the positive non trivial zero of the Yunus polynomial, which leads to
$$
e^{-\frac{\ovl{r}}{52}}=\frac{q_+}{q_++p(1-q_+)}
=\frac{1}{1+p\left(\frac{1}{q_+}-1\right)}.
$$
So $\ovl r=52\ln\left(1+p\left(\frac{1}{q_+}-1\right)\right)$.

\section{Some experimental results}

We have the chance to have with us three good students%
\footnote{L\'eo Aug\'e, Aurore Lebrun, and Ana\"\i s Pozin}
 from Polytech'Nice with 
skills in Scilab that did some numerical experiments. It turns out 
that the law for R is impressively similar to a 
Gaussian ${\cal{N}}(\mu,\sigma)$, with $\mu=\ovl r$

\begin{center}
\includegraphics[height=8cm,width=13cm]{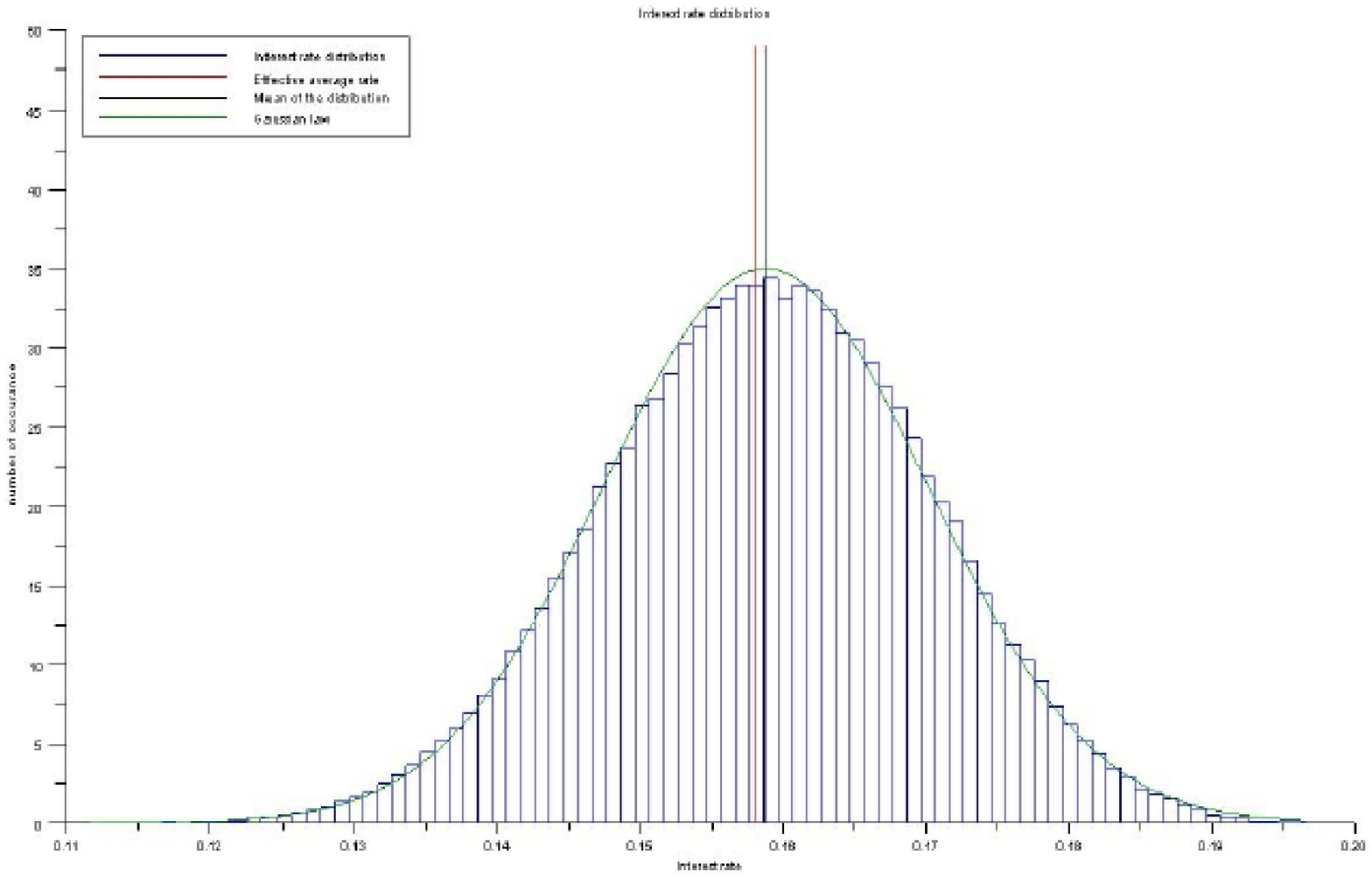}
\end{center}
On the other hand, obviously $0\leq R\leq20\%$, so $R$ can't be Gaussian, and 
indeed, even if the skewness of $R$ if very small, its kurtosis is close 
to $1$, not $3$.
\bigskip\par
\noindent So the question is (and stays, up to here)~: what is the law of $R$~?

\section{Where could infinitesimals enter the model~?}
Here some remarks related to the idea that ``50=N is large". Observe that, 
for $N=50$ and $a=10\%$ the equation ${\cal{Y}}(q)=0$ is equivalent (dividing both 
members by 20) to
\begin{equation}\label{eqphi}
\varphi(q)~:=(1+a)q^{N+1}-(N+1+a)q+N=0
\end{equation}
Now assume $N$ is infinitely large and let $q=1+\frac{x}{N}$, so $x$ is a 
blow-up of $q$ around $q=1$. Let $\psi(x):=\varphi(1+\frac{x}{N})$, so (\ref{eqphi}) 
reduces to $\psi(x)=0$. But, as $N$ is infinitely large, and denoting by 
$\zerobar$ any infinitesimal, we have
\begin{eqnarray*}
\psi(x)&=&\varphi\left(1+\frac{x}{N}\right)\\
&=&(1+a)\exp\left((N+1)\ln\left(1+\frac{x}{N}\right)\right)
     -(N+1+a)\left(1+\frac{x}{N}\right)+N \\
&=&(1+a)\exp\left((N+1)\frac{x}{N}(1+\zerobar)\right)
-x-(1+a)\left(1+\frac{x}{N}\right) \\
&=&(1+a)\exp(x+\zerobar)-x-(1+a)(1+\zerobar) \\
&\simeq&(1+a)(e^x-1)-x=:\psi_a(x).
\end{eqnarray*}
Actually, the equation $\psi_0(x)=0$ has two solutions~: $x=0$ and $-x_+<0$ so we 
get the approximation $q_+=1-\frac{x_+}{N}(1+\zerobar)$ for the non-trivial 
positive solution of ${\cal{Y}}(q)=0$. For $a=10\%$ for instance we have 
$-x_+=-0.1937476\ldots$, and $1-\frac{x_+}{50}=0.9961250\ldots$ .
\begin{figure}[h]
\begin{center}
\includegraphics[height=4.5cm,width=7cm]{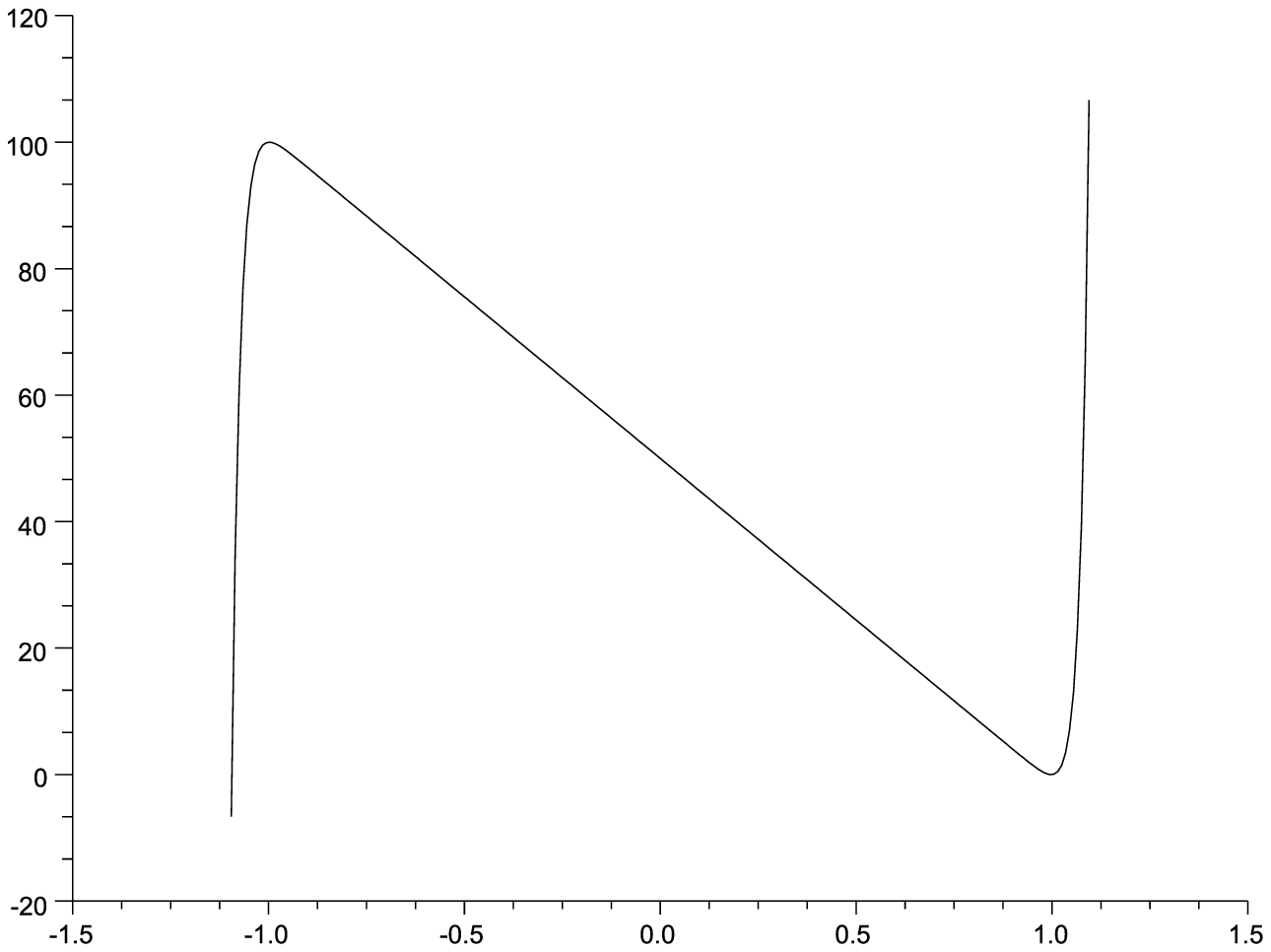}
\includegraphics[height=4.5cm,width=7cm]{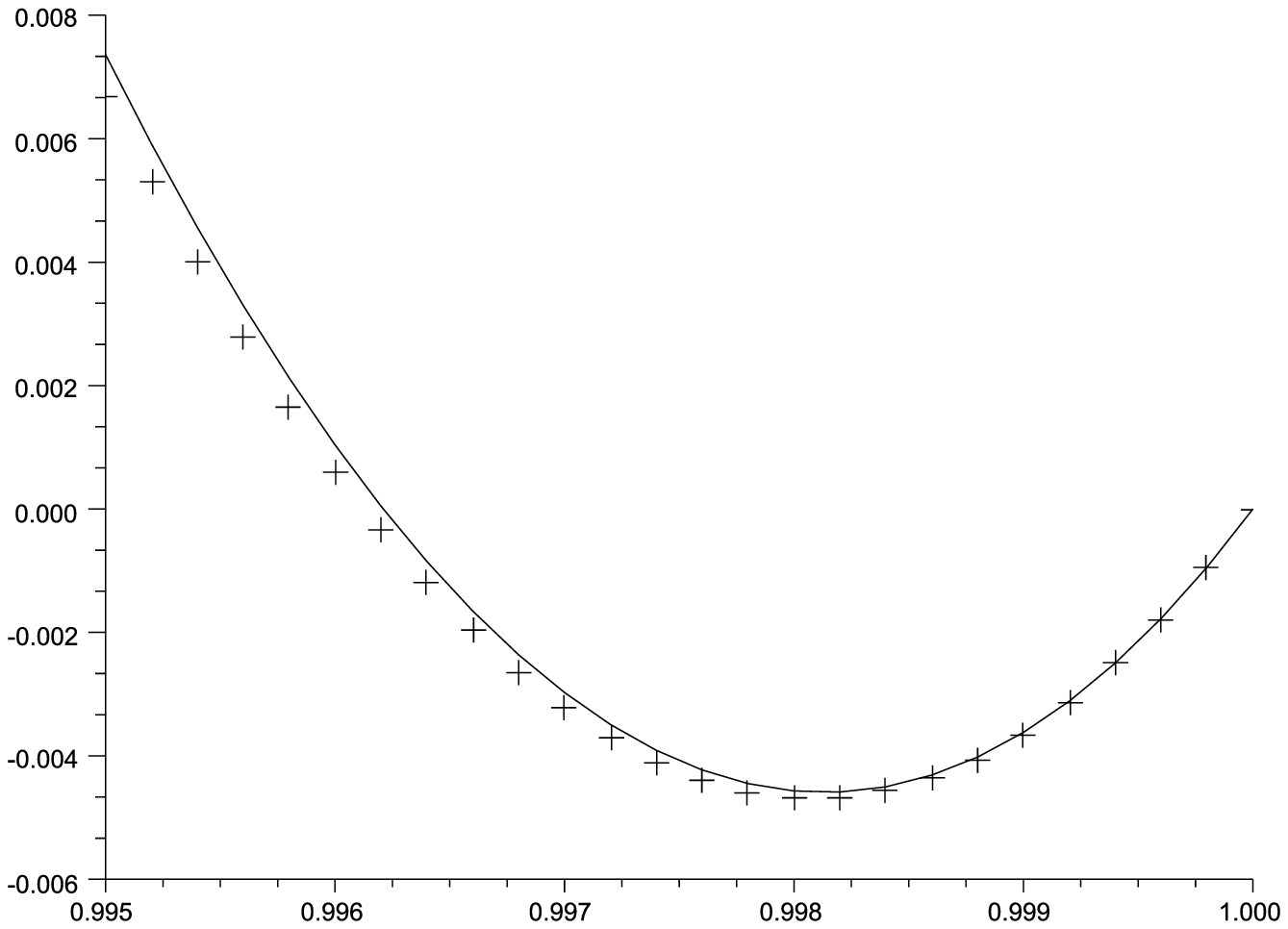}
\end{center}
\caption{The graph of function $q\mapsto\varphi(q)$, and its blow-up near $q=1$, for 
$N=50$ and $a=10\%$. The ``$+$" signs are on the graph of $x\mapsto\psi_0(x)$ rescaled so 
that $q=1+\frac{x}{N}$.}
\label{graphephipsi}
\end{figure}

So, denoting by $-x_+(a)$ the solution of $\psi_a(x)=0$ different of $1$ we have
\begin{proposition}
Assume that the number $N$ of refunds is infinitely large. Then the actuarial 
expected rate is
\begin{equation}
\ovl r(a)=\frac{1}{1+p\left(\frac{1}{1-\frac{x_+(a)}{N}(1+\zerobar)}\right)}.
\end{equation}
\end{proposition}

\bibliography{mifi}
\bibliographystyle{abbrv}
\bigskip 

\noindent 
Address of the authors:
\medskip

Universit\'e de Nice Sophia-Antipolis

Laboratoire de Math\'ematiques Jean Dieudonn\'e

Parc Valrose

06108 Nice cedex 2, France
\medskip\\
E-mail: {\tt diener@unice.fr, pheak@unice.fr}
\end{document}